\documentstyle[12pt]{article}

\def\cross{\mathord{\times}}

\begin{document}

\title{Qudit Entanglement\thanks{Work at UNM was supported in part by the
U.S.\ Office of Naval Research (Grant No.~N00014-93-1-0116), WJM acknowledges
the support of the Australian Research Council, and KN thanks the Australian
International Education Foundation (AIEF) for financial support.}}
\author{
P.~Rungta,$^{(1)}$\thanks{E-mail: pranaw@unm.edu}{\ }
W. J. Munro,$^{(2)}$\thanks{E-mail: billm@physics.uq.edu.au}{\ }
K. Nemoto,$^{(2)}$ P. Deuar,$^{(2)}$\\
G. J.  Milburn,$^{(2)}$ and C.~M. Caves$^{(1)}$\\
\\
$^{(1)}$Center for Advanced Studies,
Department of Physics and Astronomy,\\
University of New Mexico, Albuquerque, NM~87131--1156 USA\\
\\
$^{(2)}$Centre for Laser Science, Department of Physics,\\
The University of Queensland, QLD 4072 Australia
}

\date{2000 February 1}
\maketitle

\begin{abstract}

We consider the separability of various joint states of $D$-dimensional
quantum systems, which we call ``qudits.''  We derive two main results:
(i)~the separability condition for a two-qudit state that is a mixture of
the maximally mixed state and a maximally entangled state; (ii)~lower and
upper bounds on the size of the neighborhood of separable states surrounding
the maximally mixed state for $N$ qudits.

\end{abstract}

\section{Introduction}
\label{sec:intro}

One of the distinguishing features of quantum mechanics, not found in
classical physics, is the possibility of entanglement between subsystems.
It lies at the core of many applications in the emerging field of quantum
information science \cite{Lo1998}, such as quantum teleportation
\cite{Bennett1993} and quantum error correction \cite{Shor1995,Steane1996}.
Entanglement is a distinctly quantum-mechanical correlation between
subsystems, which cannot be created by actions on each subsystem separately;
moreover, correlations between subsystem measurements on an entangled
composite system cannot be explained in terms of correlations between local
classical properties inherent in the subsystems.  Thus one often says that
an entangled composite system is nonseparable.  Formally, the state of a
composite system, pure or mixed, is {\it separable\/} if the state has an
ensemble decomposition in terms of product states.  A separable state has no
quantum entanglement, and a nonseparable state is entangled.  Though the
nonclassical nature of quantum entanglement has been recognized for many
years \cite{Einstein1935,Bell1964}, only recently has considerable attention
been focused on trying to understand and characterize its properties precisely.

This paper focuses on the question of whether various joint quantum states of
$D$-di\-men\-sion\-al quantum systems are entangled.  For convenience, we call a
$D$-dimensional quantum system a ``qudit,'' by analogy with the name ``qubit''
for $D=2$ and ``qutrit'' for $D=3$.  We now have a general method for quantifying
the degree of entanglement of a pair of qubits \cite{Wootters1998}, and we have
a criterion, the partial transposition condition of Peres \cite{Peres1996},
which determines whether a general state of two qubits is entangled and whether
a general state of a qubit and a qutrit is entangled \cite{MHorodecki1996}.
The partial-transposition condition fails, however, to provide a criterion
for entanglement in other cases, where the constituents have higher
Hilbert-space dimensions \cite{PHorodecki1997,Lewenstein1999} or where there
are more than two constituents.  Indeed, at present there is no general
criterion for determining whether the joint state of $N$ qudits is entangled,
nor is there any general way to quantify the degree of entanglement if such
a state is known to be entangled.

In Sect.~\ref{sec:oprep} we review an operator representation of qudit
states, which is applied in Sect.~\ref{sec:epmax}, where we consider
states of two qudits that are a mixture of the maximally mixed state
and a maximally entangled state.  We show that such states are separable
if and only if the probability for the maximally entangled state in the
mixture does not exceed $1/(1+D)$.  This result was obtained by Horodecki
and Horodecki \cite{Horodecki1999}, and a more general result, of which
this is a special case, was obtained by Vidal and Tarrach \cite{Vidal1999}.
In Sect.~\ref{sec:prodrep} we consider the separability of mixed states of
$N$ qudits near the maximally mixed state.  We find both lower and upper
bounds on the size of the neighborhood of separable states around the
maximally mixed state.  Our results generalize and extend the results
obtained by Braunstein {\it et al.}\ for qubits \cite{Braunstein1999}
and by Caves and Milburn for qutrits \cite{Caves2000}.  Before tackling
the upper and lower bounds, we present, in Sect.~\ref{sec:math}, various
mathematical results which are used to obtain the lower bound, but which
might prove useful in other contexts as well.

\section{Operator representation of qudit states}
\label{sec:oprep}

In this section we review  an operator representation of qudit states,
analogous to the Pauli, or Bloch-sphere, representation for qubits.  We
begin with the set of Hermitian generators of SU($D$); the generators,
denoted by $\lambda_j$, are labeled by a Roman index taken from the middle
of the alphabet, which takes on values $j=1,\ldots,D^2-1$.  We represent the
generators in an orthonormal basis $|a\rangle$, labeled by a Roman letter
taken from the beginning of the alphabet, which takes on values
$a=1,\ldots,D$.  With these conventions the generators are given by
\begin{eqnarray}
&\mbox{}&
j=1,\ldots,D-1:\nonumber\\
&\mbox{}&\hphantom{j=1,\ldots,}
   \lambda_j=\Gamma_a\equiv{1\over\sqrt{a(a-1)}}
   \left(\sum_{b=1}^{a-1}
   |b\rangle\langle b|-(a-1)|a\rangle\langle a|\right)
   \;,\quad 2\leq a\leq D\;,\label{eq:diagonal}\\
&\mbox{}&
j=D,\ldots,{(D+2)(D-1)/2}:\nonumber\\
&\mbox{}&\hphantom{j=1,\ldots,}
    \lambda_j=\Gamma_{ab}^{(+)}\equiv{1\over\sqrt2}
    \left(|a\rangle\langle b|+|b\rangle\langle a|\right)
    \;,\quad 1\leq a<b\leq D\;,\label{eq:plus}\\
&\mbox{}&
j=D(D+2)/2,\ldots,D^2-1:\nonumber\\
&\mbox{}&\hphantom{j=1,\ldots,}
    \lambda_j=\Gamma_{ab}^{(-)}\equiv{-i\over\sqrt2}
    \left(|a\rangle\langle b|-|b\rangle\langle a|\right)
    \;,\quad 1\leq a<b\leq D\;.\label{eq:minus}
\end{eqnarray}
In Eqs.~(\ref{eq:plus}) and (\ref{eq:minus}), the Roman index $j$ stands for
the pair of Roman indices, $ab$, whereas in Eq.~(\ref{eq:diagonal}), it stands
for a single Roman index $a$.  The generators are traceless and satisfy
\begin{equation}
\lambda_j\lambda_k={1\over D}\delta_{jk}+d_{jkl}\lambda_l+if_{jkl}\lambda_l\;.
\end{equation}
Here and wherever it is convenient throughout this paper, we use the summation
convention to indicate a sum on repeated indices.  The coefficients $f_{jkl}$,
the structure constants of the Lie group SU($D$), are given by the commutators
of the generators and are completely antisymmetric in the three indices.  The
coefficients $d_{jkl}$ are given by the anti-commutators of the generators and
are completely symmetric.

By supplementing the $D^2-1$ generators with the operator
\begin{equation}
\lambda_0\equiv{1\over\sqrt D}I\;,
\end{equation}
where $I$ is the unit operator, we obtain a Hermitian operator basis for
the space of linear operators in the qudit Hilbert space.  This is an
orthonormal basis, satisfying
\begin{equation}
{\rm tr}
(\lambda_\alpha\lambda_\beta)=\delta_{\alpha\beta}\;.
\end{equation}
Here the Greek indices take on the values $0,\ldots,D^2-1$; throughout this
paper, Greek indices take on $D^2$ or more values.  Using this orthonormality
relation, we can invert Eqs.~(\ref{eq:diagonal})--(\ref{eq:minus}) to give
\begin{eqnarray}
|a\rangle\langle a|&=&
{I\over D}+{1\over\sqrt{a(a-1)}}
\left(-(a-1)\Gamma_a+\sum_{b=a+1}^D\Gamma_b\right)\;,
\label{eq:jkdiagonal}\\
|a\rangle\langle b|&=&
{1\over\sqrt2}(\Gamma_{ab}^{(+)}+i\Gamma_{ab}^{(-)})
\;,\quad 1\leq a<b\leq D\;,\label{eq:jkplus}\\
|b\rangle\langle a|&=&
{1\over\sqrt2}(\Gamma_{ab}^{(+)}-i\Gamma_{ab}^{(-)})
\;,\quad 1\leq a<b\leq D\;.\label{eq:jkminus}
\end{eqnarray}

Any qudit density operator can be expanded uniquely as
\begin{equation}
\label{eq:rhoex}
\rho=
{1\over D}c_\alpha\lambda_\alpha\;,
\end{equation}
where the (real) expansion coefficients are given by
\begin{equation}
c_\alpha=D{\rm tr}(\rho\lambda_\alpha)\;.
\end{equation}
Normalization implies that $c_0=\sqrt D$, so the density operator
takes the form
\begin{equation}
\label{eq:rhoexp}
 \rho={1\over D}
\left(I+c_j\lambda_j\right)
={1\over D}
(I+\vec c\cdot\vec\lambda)
\;.
\end{equation}
Here $\vec c=c_j\vec e_j$ can be regarded as a vector in a
$(D^2-1)$-dimensional real vector space, spanned by the orthonormal basis
$\vec e_j$, and $\vec\lambda=\lambda_j\vec e_j$ is an operator-valued vector.
If $\rho=|\psi\rangle\langle\psi|$ is a pure qudit state, then
${\rm tr}(\rho^2)=1$, from which it follows that
\begin{equation}
\label{eq:c}
|\vec c\,|^2=\vec c\cdot\vec c=D(D-1)\;.
\end{equation}
We could represent a pure state by a unit vector
$\vec n=\vec c/\sqrt{D(D-1)}$ on the unit sphere in $D^2-1$ dimensions, but
in contrast to the situation with the Bloch sphere ($D=2$), most vectors on
this unit sphere do not represent a pure state or, indeed, any state at all.

\section{Mixtures of maximally mixed and \\ maximally entangled states}
\label{sec:epmax}

In this section we deal with two qudits, labeled $A$ and $B$.  We consider
a class of two-qudit states, specifically mixtures of the maximally mixed
state, $M_{D^2}={I}{\otimes}{I}/{D^2}$, with a maximally entangled state,
which we can choose to be
\begin{equation}
\label{eq:maxant}
|\Psi\rangle={1\over\sqrt D}
\sum_{a=1}^{D}|a\rangle\otimes|a\rangle\;.
\end{equation}
Such mixtures have the form
\begin{equation}
\label{eq:epmaxent}
\rho_\epsilon=(1-\epsilon)M_{D^2}+\epsilon|\Psi\rangle\langle\Psi|\;,
\end{equation}
where $0\le\epsilon\le1$.

In analogy to Eq. (\ref{eq:rhoex}), any state $\rho$ of two qudits can be
expanded uniquely as
\begin{equation}
\rho=
{1\over D^2}c_{\alpha\beta}\lambda_\alpha\otimes\lambda_\beta\;,
\end{equation}
where the expansion coefficients are given by
\begin{equation}
\label{eq:expco}
c_{\alpha\beta}=D^2{\rm tr}
(\rho\lambda_\alpha\otimes\lambda_\beta)\;,
\end{equation}
with $c_{00}=D$ determined by normalization.  Using
Eq.~(\ref{eq:expco}) or Eqs.~(\ref{eq:jkdiagonal})--(\ref{eq:jkminus}), we can
find the operator expansion for the maximally entangled state~(\ref{eq:maxant}):
\begin{equation}
|\Psi\rangle\langle\Psi|
={1\over D^2}\!\left(
I\otimes I
+D\sum_a\Gamma_a\otimes\Gamma_a
+D\sum_{a<b}\left(\Gamma_{ab}^{(+)}\otimes\Gamma_{ab}^{(+)}
-\Gamma_{ab}^{(-)}\otimes\Gamma_{ab}^{(-)}\right)\right)\;,
\end{equation}
from which we can read off the expansion coefficients for the state
$\rho_{\epsilon}$ of Eq.~(\ref{eq:epmaxent}):
\begin{eqnarray}
c_{0j}&=&c_{j0}=0\;,\\
c_{jk}&=&\cases{
0\;,&$j\ne k$,\cr
D\epsilon\;,&$j=k=1,\ldots,(D+2)(D-1)/2$.\cr
-D\epsilon\;,&$j=k=D(D+2)/2,\ldots,D^2-1$.}
\label{eq:coeff}
\end{eqnarray}

A state of the two qudits is {\it separable\/} if it can be written as an
ensemble of product states.  In this section we show that the mixed state
state~(\ref{eq:epmaxent}) is separable if and only if
\begin{equation}
\epsilon\le{1\over1+D}\;.
\label{eq:sepboundary}
\end{equation}
Our method is to prove the necessity of the condition~(\ref{eq:sepboundary})
by considering the restrictions that separability places on the correlation
coefficients~(\ref{eq:coeff}) and then to construct an explicit product
ensemble when $\epsilon\le1/(1+D)$.  Vidal and Tarrach \cite{Vidal1999}
found the separability boundary for a mixture of $M_{D^2}$ with {\it any\/}
pure state by using the partial transpose condition~\cite{Peres1996} to show
that any state with $\epsilon$ outside the boundary is nonseparable and by
constructing an explicit product ensemble for states with $\epsilon$ within
the separability boundary.  Horodecki and Horodecki \cite{Horodecki1999}
found the separability boundary for the state~(\ref{eq:epmaxent}) using
other techniques.  The reason for presenting in this section a more limited
result than that of Vidal and Tarrach is, first, that our proof of necessity
has a nice physical interpretation in terms of the correlation
coefficients~(\ref{eq:coeff}) and, second, that the product
ensemble we use is different from the one used by Vidal and Tarrach.

The product pure states for two qudits,
$|\psi_A\rangle\langle\psi_A|{\otimes}|\psi_B\rangle\langle\psi_B|$,
constitute an overcomplete operator basis. Thus we can expand any
two-qudit density operator in terms of them,
\begin{equation}
\rho=\int {d{\cal V}}_ {A}\,{d{\cal
V}}_{B}\,{w({\psi_A},{\psi_B})}\,|\psi_A\rangle\langle\psi_A|
{\otimes}|\psi_B\rangle\langle\psi_B|\;.
\end{equation}
Here the integral for each system runs over all of projective Hilbert space,
i.e., the space of Hilbert-space rays, and the volume elements $d{\cal V}_A$
and $d{\cal V}_B$ are the unitarily invariant integration measures on projective
Hilbert space.  Because of overcompleteness of the pure-state projectors, the
expansion function ${w({\psi_A},{\psi_B})}$ is not unique.  Notice that the
expansion coefficients $c_{{\alpha}{\beta}}$ of Eq.~(\ref{eq:expco}) can be
written as integrals over the expansion function,
\begin{equation}
\label{eq:exco}
 c_{{\alpha}{\beta}}=\int {d{\cal V}}_ {A}\,{d{\cal
V}}_{B} \,{w({\psi_A},{\psi_B})}\, (c_A)_{\alpha}(c_B)_{\beta}\;,
\end{equation}
where $(c_A)_\alpha=D\langle\psi_A|\lambda_\alpha|\psi_A\rangle$ and
$(c_B)_\alpha=D\langle\psi_B|\lambda_\alpha|\psi_B\rangle$ are the expansion
coefficients for the pure states $|\psi\rangle_A$ and $|\psi\rangle_B$,
satisfying $\vec c_A\cdot\vec c_A=D(D-1)=\vec c_B\cdot\vec c_B$.

A two-qudit state is separable if and only if there exists an expansion
function $w(\psi_A,\psi_B)$ that is everywhere nonnegative.  In this case
$w(\psi_A,\psi_B)$ can be thought of as a normalized classical probability
distribution for the pure states $\psi_A$ and $\psi_B$, and the integral
for $c_{\alpha\beta}$ in Eq.~(\ref{eq:exco}) can be interpreted as a classical
expectation value of the product of the random variables $(c_A)_\alpha$
and $(c_B)_\beta$, i.e.,
\begin{equation}
\label{eq:coex}
c_{\alpha\beta}=E\left[(c_A)_\alpha(c_B)_\beta\right]\;.
\end{equation}
If the state $\rho_\epsilon$ is separable, we have from Eq.~(\ref{eq:coeff})
that for each value of $j$,
\begin{equation}
D{\epsilon}=|c_{jj}|
=\Bigl|E\left[(c_A)_j(c_B)_j\right]\Bigr|
\leq
{1\over2}\!\left(E[(c_A)_j^2]+E[(c_B)_j^2]\right)\;.
\end{equation}
Adding over the $D^2-1$ value of $j$ gives
\begin{equation}
D(D^2-1)\epsilon
\leq
{1\over2}
\Bigl(
E[\vec c_A\cdot\vec c_A]+E[\vec c_B\cdot\vec c_B]
\Bigr)
=D(D-1)\;.
\end{equation}
We conclude that if $\rho_\epsilon$ is separable, then $\epsilon\le1/(1+D)$.

To prove the converse, we construct an explicit product ensemble for the
state $\rho_\epsilon$ with $\epsilon=1/(1+D)$.  We define a vector
$\vec z=(z_1,\ldots,z_D)$ whose components $z_a$ take on the values
$\pm1$ and $\pm i$, so that
\begin{equation}
\label{eq:z} {\sum_{z_j}}{z_j}={\sum_{z_j}}{z_j^2}=0\;, \, \,
\sum_{z_j}|z_j|^2=4\;.
\end{equation}
Associated with each vector $\vec z$ is a pure state
\begin{equation}
\label{eq:maxanty}
|\Phi_{\vec z}\rangle={1\over\sqrt D}\sum_{a=1}^{D}z_a|a\rangle\;.
\end{equation}
There are $4^D$ vectors and thus that many states $\Phi_{\vec z}$, although
only $4^{D-1}$ of these states are distinct in that they differ by more than
a global phase.  Now we define a product state for the two-qudit system:
\begin{equation}
\label{eq:rho}
\rho_{\vec z}=|\Phi_{\vec z}\rangle\langle\Phi_{\vec z}|
\otimes|
\Phi_{\vec z^*}\rangle\langle\Phi_{\vec z^*}|\;.
\end{equation}
The ensemble consisting of all $4^D$ of these states, each contributing with
the same probability, produces the density operator
\begin{equation}
{1\over4^D}\sum_{\vec z}\rho_{\vec z}=
{1\over4^D D^2}\sum_{a,b,c,d}
\left(\sum_{\vec z}z_a z_b^\ast z_c^\ast z_d\right)
|a\rangle\langle b|\otimes|c\rangle\langle d|\;.
\end{equation}
Since
\begin{equation}
\sum_{\vec z}z_a z_b^\ast z_c^\ast z_d=
4^D(\delta_{ab}\delta_{cd}+\delta_{ac}\delta_{bd}-
\delta_{ab}\delta_{cd}\delta_{ac})\;,
\end{equation}
it follows that
\begin{equation}
{1\over4^D}\sum_{\vec z}\rho_{\vec z}=
{I\otimes I\over D^2}
+{1\over D}|\Psi\rangle\langle\Psi|-
{1\over D^2}\sum_{a=1}^D
|a\rangle\langle a|\otimes|a\rangle\langle a|\;.
\end{equation}
Multiplying by $D/(D+1)$ and rearranging yields
\begin{equation}
{D\over1+D}{I\otimes I\over D^2}+
{1\over 1+D}|\Psi\rangle\langle\Psi|
={D\over1+D}{1\over4^D}
\sum_{\vec z}\rho_{\vec z}+
{1\over1+D}{1\over D}\sum_{a=1}^D
|a\rangle \langle a|\otimes|a\rangle\langle a|\;.
\label{eq:rhom}
\end{equation}
The left-hand side of Eq.~(\ref{eq:rhom}) is the state~(\ref{eq:epmaxent})
with $\epsilon=1/(1+D)$, and the right-hand side is an explicit product ensemble
for the state.  This concludes the proof that $\rho_\epsilon$ is separable if
and only if $\epsilon\leq1/(1+D)$.

\section{Separability of states near the\\maximally mixed state}
\label{sec:prodrep}

This section deals with $N$-qudit states of the form
\begin{equation}
\rho_\epsilon=(1-\epsilon)M_{D^N}+\epsilon\rho_1\ \;,
\label{eq:rhoep}
\end{equation}
where $M_{D^N}=I\otimes\cdots\otimes I/D^N$ is the maximally mixed state
for $N$ qudits and $\rho_1$ is any $N$-qudit density operator.  We establish
lower and upper bounds on the size of the neighborhood of separable states
surrounding the maximally mixed state. In particular, we show, first, that for
\begin{equation}
\epsilon\le{1\over1+D^{2N-1}}\;,
\label{eq:lbound}
\end{equation}
all states of the form~(\ref{eq:rhoep}) are separable and, second, that for
\begin{equation}
\epsilon>{1\over1+D^{N-1}}\;,
\label{eq:ubound}
\end{equation}
there are states of the form~(\ref{eq:rhoep}) that are not separable (i.e.,
they are entangled).  These results generalize and extend the work of
Braunstein {\it et al.}\ for qubits \cite{Braunstein1999} and of Caves and
Milburn for qutrits \cite{Caves2000}.

\subsection{Mathematical preliminaries}
\label{sec:math}

Before turning to the lower and upper bounds, it is useful to develop some
mathematical apparatus that will be used in deriving the bounds.

\subsubsection{Superoperator formalism}
\label{sec:super}

We begin by reviewing a formalism for handling superoperators, introduced
by Caves \cite{Caves1999} and used by Schack and Caves \cite{Schack2000}
to generate product ensembles for separable $N$-qubit states.

The space of linear operators acting on a $D$-dimensional complex vector
space is a $D^2$-dimensional complex vector space.  In this space we
introduce operator ``kets'' $|A)=A$ and ``bras'' $(A|=A^\dagger$,
distinguished from vector kets and bras by the use of smooth brackets.
The natural operator inner product can be written as
$(A|B)={\rm tr}(A^\dagger B)$.  An orthonormal basis $|a\rangle$ induces
an orthonormal operator basis,
\begin{equation}
|c\rangle\langle a|=\tau_{ca}=\tau_\alpha\;,
\end{equation}
where the Greek index $\alpha$ is an abbreviation for the pair of Roman
indices, $ca$.  Not all orthonormal operator bases are of this outer-product
form.

The space of superoperators, i.e., linear maps on operators, is a
$D^4$-dimensional complex vector space.  Any superoperator ${\cal S}$
is specified by its ``matrix elements''
\begin{equation}
{\cal S}_{ca,db}=
\langle c|\,{\cal S}(|a\rangle\langle b|)|d\rangle\;,
\label{eq:matel1}
\end{equation}
for the superoperator can be written in terms of its matrix
elements as
\begin{equation}
{\cal S}=\sum_{c,a,d,b} {\cal S}_{ca,db}
|c\rangle\langle a|\odot|b\rangle\langle d|
=\sum_{c,a,d,b}
{\cal S}_{ca,db}\,\tau_{ca}\odot\tau_{db}^\dagger=
\sum_{\alpha,\beta} {\cal S}_{\alpha\beta}|\tau_\alpha)(\tau_\beta|\;.
\label{eq:S}
\end{equation}
The tensor product here is an ordinary operator tensor product, but we
use the symbol $\odot$ to distinguish it from a tensor product between
objects associated with different systems, which is denoted by $\otimes$.
In the final form of Eq.~(\ref{eq:S}), the tensor product is written as an
operator outer product, with $\alpha=ca$ and $\beta=db$.

The ordinary action of ${\cal S}$ on an operator $A$, used to generate
the matrix elements, is obtained by dropping an operator $A$ into the center
of the representation of ${\cal S}$, in place of the tensor-product sign,
\begin{equation}
{\cal S}(A)= \sum_{\alpha,\beta} {\cal
S}_{\alpha\beta}\,\tau_\alpha A\tau_\beta^\dagger\;.
\end{equation}
There is clearly another way that ${\cal S}$ can act on $A$, the left-right
action,
\begin{equation}
{\cal S}|A)=\sum_{\alpha,\beta} {\cal S}_{\alpha\beta}
|\tau_\alpha)(\tau_\beta|A)\;,
\end{equation}
in terms of which the matrix elements are
\begin{equation}
{\cal S}_{\alpha\beta}
=(\tau_\alpha|\,{\cal S}|\tau_\beta)
=(\tau_{ca}|\,{\cal S}|\tau_{db})
=\langle c|\,{\cal S}(|a\rangle\langle b|)|d\rangle \;.
\label{eq:matel2}
\end{equation}
This expression provides the fundamental connection between the two actions
of a superoperator.  We can define an operation, called {\it sharp}, that
exchanges the ordinary and left-right actions:
\begin{equation}
{\cal S}^{\#}(A)={\cal S}|A)\;.
\end{equation}
Equation~(\ref{eq:matel2}) implies that
\begin{equation}
{\cal S}^{\#}_{ca,db}=
\langle c|\,{\cal S}^{\#}(|a\rangle\langle b|)|d\rangle
=(\tau_{cd}|{\cal S}|\tau_{ab})=
{\cal S}_{cd,ab}
\end{equation}
or, equivalently, that
\begin{equation}
{\cal S}^{\#}=\sum_{c,a,d,b} {\cal S}_{ca,db}
|c\rangle\langle d|\odot|b\rangle\langle a|\;.
\end{equation}

With respect to the left-right action, a superoperator works just
like an operator.  Multiplication of superoperators ${\cal R}$ and
${\cal S}$ is given by
\begin{equation}
{\cal R\cal S}= \sum_{\alpha,\beta,\gamma}
{\cal R}_{\alpha\gamma}{\cal S}_{\gamma\beta}
|\tau_\alpha)(\tau_\beta|\;,
\end{equation}
and the adjoint is defined by
\begin{equation}
(A|{\cal S}^\dagger|B)=(B|{\cal S}|A)^*
\quad\Longleftrightarrow\quad {\cal S}^\dagger=
\sum_{\alpha,\beta} {\cal S}_{\beta\alpha}^*
|\tau_\alpha)(\tau_\beta|\;.
\end{equation}
With respect to the ordinary action, superoperator multiplication,
denoted as a composition ${\cal R}\circ{\cal S}$, is given by
\begin{equation}
{\cal R}\circ{\cal S}= \sum_{\alpha,\beta,\gamma,\delta}
{\cal R}_{\gamma\delta}{\cal S}_{\alpha\beta}\,\tau_\gamma\tau_\alpha
\odot\tau_\beta^\dagger\tau_\delta^\dagger\;.
\end{equation}
The adjoint with respect to the ordinary action, denoted by ${\cal S}^{\cross}$,
is defined by
\begin{equation}
{\rm tr}\bigl([{\cal S}^{\cross}(B)]^\dagger A\bigr)=
{\rm tr}\bigl(B^\dagger{\cal S}(A)\bigr) \quad\Longleftrightarrow\quad
{\cal S}^{\cross}=\sum_{\alpha,\beta} {\cal S}_{\alpha\beta}^*\,
\tau_\alpha^\dagger\odot\tau_\beta\;.
\end{equation}

The identity superoperator with respect to the left-right action can be
written as
\begin{equation}
{\bf I}=\sum_\alpha|\tau_\alpha)(\tau_\alpha|
=\sum_{c,a}|c\rangle\langle a|\odot|a\rangle\langle c|\;.
\end{equation}
When sharped, ${\bf I}$ becomes the identity superoperator with respect
to the ordinary action, denoted by ${\cal I}$:
\begin{equation}
{\bf I}^\#=\sum_{c,a}|c\rangle\langle c|\odot|a\rangle\langle a|
=I\odot I\equiv{\cal I}\;.
\end{equation}
The final ingredient we need is the superoperator trace relative to
the left-right action, defined by
\begin{equation}
{\rm Tr}({\cal S})=\sum_\alpha(\tau_\alpha|{\cal S}|\tau_\alpha)=
\sum_{c,a}\langle c|\,{\cal S}(|a\rangle\langle a|)|c\rangle
={\rm tr}({\cal S}(I))\;.
\end{equation}
Notice that ${\bf I}(I)=DI$ and ${\cal I}(I)=I$, which give
${\rm Tr}({\bf I})=D^2$ and ${\rm Tr}({\cal I})=D$.

Now suppose the operators $|N_\alpha)$ constitute a complete or overcomplete
operator basis; i.e., let the operator kets $|N_\alpha)$ span the vector space
of operators.  It follows that the superoperator ${\cal G}$ defined by
\begin{equation}
\label{eq:g}
{\cal G}=\sum_\alpha|N_\alpha)(N_\alpha|={\cal G}^\dagger
\end{equation}
is invertible with respect to the left-right action.  The operators
\begin{equation}
\label{eq:d}
|Q_\alpha)={\cal G}^{-1}|N_\alpha)
\end{equation}
form a dual basis, which gives rise to the following expressions for the
identity superoperator:
\begin{equation}
\label{eq:resid}
{\bf I}=\sum_\alpha|Q_\alpha)(N_\alpha|=\sum_\alpha|N_\alpha)(Q_\alpha|
\;.
\end{equation}
An arbitrary operator $A$ can be expanded in terms of the original basis
or the dual basis:
\begin{eqnarray}
\label{eq:expansion1}
A&=&\sum_\alpha|N_\alpha)(Q_\alpha|A)=
\sum_\alpha N_\alpha{\rm tr}(Q_\alpha^\dagger A)\;,\\
\label{eq:expansion2}
A&=&\sum_\alpha|Q_\alpha)(N_\alpha|A)
=\sum_\alpha Q_\alpha{\rm tr}(N_\alpha^\dagger A)\;.
\end{eqnarray}
These expansions are unique if and only if the operators $|N_\alpha)$ are
linearly independent.  Later in this section we apply expansions of this
sort to density operators.

\subsubsection{Pure states and their dual basis}
\label{sec:dualb}

The set of all pure-state projectors in a $D$-dimensional Hilbert space,
\begin{equation}
P_\psi=|\psi\rangle\langle\psi|\;,
\end{equation}
forms an overcomplete operator basis.  To develop operator expansions in terms
of the pure-state projectors, we follow the discussion in the preceding
subsection and consider the superoperator
\begin{equation}
\label{eq:intG}
{\cal G}=\int d{\cal V}\,|P_\psi)(P_\psi|
=\int d{\cal V}\,|\psi\rangle\langle\psi|\odot|\psi\rangle\langle\psi|\;,
\end{equation}
where $d{\cal V}$ is the unitarily invariant integration measure on projective
Hilbert space.

The only Hilbert-space integrals we need to calculate explicitly are those
for which the integrand is a function only of an angle $\theta$ defined
by $\cos\theta=|\langle e|\psi\rangle|$, where $|e\rangle$ is some
particular unit vector (pure state).  The angle $\theta$, which runs over
the range $0\le\theta\le\pi/2$, can be thought of as a ``polar angle''
relative to the ``polar axis'' defined by $|e\rangle$. For integrals of this
sort, a convenient form of the integration measure is \cite{Schack1994}
\begin{equation}
d{\cal V}=(\sin\theta)^{2D-3}\cos\theta\,d\theta\,d{\cal S}_{2D-3}\;,
\end{equation}
where $d{\cal S}_{2D-3}$ is the standard integration measure on a
$(2D-3)$-dimensional unit sphere.  Thus the total volume of $D$-dimensional
projective Hilbert space is \cite{Schack1994}
\begin{equation}
\label{eq:vol}
{\cal V}=
{\cal S}_{2D-3}
\int_0^{\pi/2}d\theta\,(\sin\theta)^{2D-3}\cos\theta\,d\theta
={{\cal S}_{2D-3}\over2(D-1)}
={\pi^{D-1}\over(D-1)!}\;,
\end{equation}
where ${\cal S}_{2D-3}=2\pi^{D-1}/(D-2)!$ is the volume of a
$(2D-3)$-dimensional unit sphere.

To use the expansions~(\ref{eq:expansion1}) and (\ref{eq:expansion2}),
we need the dual basis $|Q_\psi)$, and for that purpose, we need to invert
${\cal G}$.  Since ${\cal G}$ is Hermitian relative to the left-right
action, we can invert it by diagonalizing it with respect to the left-right
action.  Given an orthonormal basis $|a\rangle$, we can write ${\cal G}$
as in Eq.~(\ref{eq:S}),
\begin{equation}
{\cal G}
=\sum_{c,a,d,b} {\cal G}_{ca,db} |c\rangle\langle a|\odot|b\rangle\langle d|
=\sum_{c,a,d,b} {\cal G}_{ca,db} |\tau_{ca})(\tau_{db}|\;,
\end{equation}
where the matrix elements are given by Eq.~(\ref{eq:matel1}):
\begin{equation}
{\cal G}_{ca,db}=
\langle c|\,{\cal G}(|a\rangle\langle b|)|d\rangle=
\int d{\cal V}\,
\langle c|\psi\rangle
\langle\psi|a\rangle
\langle b|\psi\rangle
\langle\psi|d\rangle\;.
\label{eq:Gmatel}
\end{equation}

The unitary invariance of the integration measure places stringent constraints
on the matrix elements~(\ref{eq:Gmatel}).  Since the integral in
Eq.~(\ref{eq:Gmatel}) remains unchanged under a change in the sign of the
amplitude $\langle a|\psi\rangle$ corresponding to a particular basis vector
$|a\rangle$, the matrix elements vanish except when (i)~$a=b\ne c=d$ or
$a=c\ne b=d$ or (ii)~$a=b=c=d$.  Furthermore, unitary invariance implies that
for each of these cases, all the matrix elements have the same value.
Gathering these conclusions together, we have
\begin{equation}
\label{eq:matele}
{\cal G}_{ca,db}=\cases{
\alpha\;,&$a=b\ne c=d$ or $a=c\ne b=d$,\cr
\gamma\;,&$a=b=c=d$,\cr
0\;,&otherwise.}
\end{equation}

We get a relation between $\alpha$ and $\gamma$ by noting that
\begin{equation}
\label{eq:mate}
D(D-1)\alpha+D\gamma
=\sum_{c,a=1}^D{\cal G}_{ca,ca}={\cal V}\;,
\end{equation}
where the second equality follows from doing the sum within the integral
in Eq.~(\ref{eq:Gmatel}).  We need one more relation, which we get by
evaluating explicitly the integral for $\gamma$:
\begin{equation}
\label{eq:mateq3}
\gamma=\int d{\cal V}|\langle a|\psi\rangle|^4
={\cal S}_{2D-3}\int_0^{\pi/2}
d\theta\,(\sin\theta)^{2D-3}(\cos\theta)^5
={2{\cal V}\over D(D+1)}\equiv 2K\;.
\end{equation}
It follows that $\alpha=K$.  As a result, we have
\begin{equation}
\label{eq:G}
{\cal G}=K\Biggl(
2\sum_a|\tau_{aa})(\tau_{aa}|+
\sum_{{\scriptstyle{a,b}\atop\scriptstyle{a\ne b}}}
\Bigl(|\tau_{ab})(\tau_{ab}|+ |\tau_{aa})(\tau_{bb}|\Bigl)\Biggr)
=K({\bf I}+{\cal I})\;.
\end{equation}
This result gives us immediately that \cite{Schack1994}
\begin{equation}
\int d{\cal V}\,|\psi\rangle\langle\psi|=
{\cal G}(I)={{\cal V}\over D}I\;.
\end{equation}

The operators $\lambda_\alpha$ introduced in Sect.~\ref{sec:oprep} constitute
a complete, orthonormal operator basis, so we can write ${\bf I}$ as
\begin{equation}
{\bf I}=\sum_\alpha|\lambda_\alpha)(\lambda_\alpha|
={|I)(I|\over D}+{\cal T}\;,
\label{eq:Ilambda}
\end{equation}
where
\begin{equation}
{\cal T}=\sum_j|\lambda_j)(\lambda_j|
\end{equation}
is the superoperator that projects onto the subspace of traceless operators.
Plugging Eq.~(\ref{eq:Ilambda}) into Eq.~(\ref{eq:G}) gives the diagonal
form of ${\cal G}$:
\begin{equation}
{\cal G}=K\!\left((D+1){|I)(I|\over D}+{\cal T}\right)\;.
\label{eq:Gdiagonal}
\end{equation}
Orthonormal eigenoperators of ${\cal G}$ are $\lambda_0=I/\sqrt D$,
with eigenvalue $K(D+1)={\cal V}/D$ and the traceless operators
$\lambda_j$, which are degenerate with eigenvalue $K={\cal V}/D(D+1)$.

We are now prepared to write the inverse of ${\cal G}$ with respect to the
left-right action as
\begin{equation}
{\cal G}^{-1}={1\over K}\!\left({1\over D+1}{|I)(I|\over D}+{\cal T}\right)
={1\over K}\!\left({\bf I}-{{\cal I}\over D+1}\right)\;.
\end{equation}
Thus the dual operators of Eq.~(\ref{eq:d}) are given by
\begin{equation}
\label{eq:gri}
|Q_\psi)={\cal G}^{-1}|P_\psi)
={1\over K}\!\left(|P_\psi)-{|I)\over D+1}\right)
={D\over{\cal V}}\Bigl((D+1)P_\psi-I\Bigr)
\;.
\end{equation}

\subsubsection{Alternative diagonalization of ${\cal G}$}
\label{sec:ad}

In this subsection we rederive Eq.~(\ref{eq:Gdiagonal}) using the special
properties of the superoperator ${\cal G}$.  These properties are evident
from the integral form of ${\cal G}$ in Eq.~(\ref{eq:intG}).

\begin{itemize}

\begin{item}
The superoperator ${\cal G}$ is Hermitian relative to the left-right
action, which implies that it has a complete, orthonormal set of eigenoperators
$\eta_\alpha$, $\alpha=1,\ldots,D^2$, with real eigenvalues $q_\alpha$:
\begin{equation}
{\cal G}={\cal G}^{\dagger}
\quad\Longrightarrow\quad
{\cal G}=\sum_\alpha q_\alpha|\eta_\alpha)(\eta_\alpha|=
\sum_\alpha q_\alpha\eta_\alpha\odot\eta_\alpha^\dagger
\;.
\end{equation}
\end{item}

\begin{item}
The superoperator ${\cal G}$ is Hermitian relative to the ordinary action,
\begin{equation}
{\cal G}={\cal G}^{\cross}
=\sum_\alpha q_\alpha\eta_\alpha^\dagger\odot\eta_\alpha
=\sum_\alpha q_\alpha|\eta_\alpha^\dagger)(\eta_\alpha^\dagger|
\;,
\end{equation}
which implies that if $\eta_\alpha$ is an eigenoperator of ${\cal G}$,
then $\eta_\alpha^\dagger$ is also an eigenoperator with the same eigenvalue.
This means that we can choose all the eigenoperators to be Hermitian.
\end{item}

\begin{item}
The superoperator ${\cal G}$ is unitarily invariant, i.e.,
\begin{equation}
{\cal G}=U\odot U^\dagger\circ{\cal G}\circ U^\dagger\odot U
=\sum_\alpha q_\alpha U\eta_\alpha U^\dagger\odot U\eta_\alpha^\dagger U^\dagger
\;,
\end{equation}
for any unitary operator $U$, which implies that if $\eta_\alpha$ is an
eigenoperator of ${\cal G}$, then $U\eta_\alpha U^\dagger$ is also
an eigenoperator with the same eigenvalue.
\end{item}
\end{itemize}

The upshot of these three properties is that the eigensubspaces of
${\cal G}$ are invariant under Hermitian conjugation and under all
unitary transformations.  It is not hard to show that the only such
operator subspaces are the subspace of traceless operators and its
orthocomplement, the one-dimensional subspace spanned by the unit operator.
The result is that ${\cal G}$ must have the form
\begin{equation}
{\cal G}
=K\!\left(\mu{I\odot I\over D}+{\cal T}\right)
=K\!\left({\bf I}+{\mu-1\over D}{\cal I}\right)
\;,
\end{equation}
where $K$ is the eigenvalue of any traceless operator and $K\mu$ is the
eigenvalue of $\lambda_0=I/\sqrt D$.  Now we use the final property to
evaluate $\mu$.

\begin{itemize}
\begin{item}
The superoperator ${\cal G}$ is invariant under exchange of the two kinds
of action:
\begin{equation}
{\cal G}={\cal G}^{\#}
=K\!\left({\cal I}+{\mu-1\over D}{\bf I}\right)\;.
\end{equation}
This implies that $\mu=D+1$, thus bringing ${\cal G}$ into the form~(\ref{eq:G}),
but with $K$ not yet determined.
\end{item}
\end{itemize}

\noindent
We find the value of $K$ by evaluating the superoperator trace, first using
Eq.~(\ref{eq:intG}),
\begin{equation}
{\rm Tr}({\cal G})={\rm tr}({\cal G}(I))={\cal V}\;,
\end{equation}
and then using Eq.~(\ref{eq:G}),
\begin{equation}
{\rm Tr}({\cal G})=
K\Bigl({\rm Tr}({\bf I})+{\rm Tr}({\cal I})\Bigr)=KD(D+1)\;.
\end{equation}
This gives $K={\cal V}/D(D+1)$, in agreement with Eq.~(\ref{eq:mateq3}).

\subsection{Separability bounds}

We turn now to demonstrating the lower and upper bounds, Eqs.~(\ref{eq:lbound})
and (\ref{eq:ubound}), on the size of the neighborhood of separable states
surrounding the maximally state.

To establish the lower bound, we use the results of Sect.~\ref{sec:math}
to formulate operator expansions in terms of product pure states.  For a
single qudit, any density operator can be expanded as
\begin{equation}
\rho=\int d{\cal V}\,|P_\psi)(Q_\psi|\rho)=
\int d{\cal V}\,w_\rho(\psi)P_\psi\;,
\end{equation}
where
\begin{equation}
w_\rho(\psi)={\rm tr}(\rho Q_\psi)
={D\over{\cal V}}\Bigl((D+1)\langle\psi|\rho|\psi\rangle-1\Bigr)
\end{equation}
is a quasi-probability distribution, normalized to unity, but possibly
having negative values.  The analogous product representation for an
$N$-qudit density operator is
\begin{equation}
\label{eq:r}
\rho=\int d{\cal V}_1\cdots d{\cal V}_N\,
w_\rho(\psi_1,\ldots,\psi_N)\,
P_{\psi_1}\otimes\cdots\otimes P_{\psi_N}\;,
\end{equation}
where
\begin{equation}
\label{eq:w}
w_\rho(\psi_1,\ldots,\psi_N)=
{\rm tr}(\rho Q_{\psi_1}\otimes\cdots\otimes Q_{\psi_N})\;.
\end{equation}
The $N$-qudit quasi-distribution obeys the bound
\begin{equation}
\label{eq:wb}
w_\rho(\psi_1,\ldots,\psi_N)\ge
\pmatrix{
\hbox{smallest eigenvalue of}\cr
Q_{\psi_1}\otimes\cdots\otimes Q_{\psi_N}}
=-{D^{2N-1}\over{\cal V}^N}
\end{equation}
This follows from the fact that $Q_\psi$ has a nondegenerate eigenvalue,
$D^2/{\cal V}$, and a $(D-1)$-fold degenerate eigenvalue, $-D/{\cal V}$.
Thus the most negative eigenvalue of the product operator
$Q_{\psi_1}\otimes\cdots\otimes Q_{\psi_N}$ is
$(-D/{\cal V})(D^2/{\cal V})^{N-1}=-D^{2N-1}/{\cal V}^N$.

We can use the lower bound~(\ref{eq:wb}) to place a similar lower bound on
the quasi-distribution for the mixed state $\rho_\epsilon$ of
Eq.~(\ref{eq:rhoep}).  Since the quasi-distribution for the maximally
mixed state, $M_{D^N}$, is the uniform distribution $1/{\cal V}^N$, we have
\begin{equation}
w_{\rho_\epsilon}(\psi_1,\ldots,\psi_N)
={1-\epsilon\over{\cal V}^N}
+\epsilon w_{\rho_1}
\ge{1-\epsilon(1+D^{2N-1})\over{\cal V}^N}\;.
\end{equation}
We conclude that if $\epsilon\le1/(1+D^{2N-1})$, then $w_{\rho_{\epsilon}}$
is nonnegative and the qudit state $\rho_{\epsilon}$ is separable.  This
establishes the lower bound~(\ref{eq:lbound}) on the size of the neighborhood
of separable states surrounding the maximally mixed state.

The upper bound~(\ref{eq:ubound}) on the size of the separable neighborhood
can be established with the help of an exact separability condition for a
particular $N$-qubit state, obtained by D{\"u}r, Cirac, and Tarrach
\cite{Dur1999} and also by Pittenger and Rubin \cite{Pittenger1999}.
We consider the $N$-qudit state,
\begin{equation}
\label{eq:mcat}
\rho_{\epsilon}=
(1-\epsilon)M_{D^N}+\epsilon|\Psi_{\rm cat}\rangle\langle\Psi_{\rm cat}|\;,
\end{equation}
where
\begin{equation}
\label{eq:cat}
 |\Psi_{\rm cat}\rangle=
{1\over\sqrt D}\sum_{a=1}^D
|a\rangle\otimes\cdots\otimes|a\rangle\;,
\end{equation}
is an $N$-qudit ``cat state.''  We call the mixed state~(\ref{eq:mcat})
an $\epsilon$-cat state.

Now project each qudit onto the two-dimensional (qubit) subspace spanned by
$|1\rangle$ and $|2\rangle$.  The local projection operator on each qudit
is $\Pi=|1\rangle\langle1|+|2\rangle\langle2|$, and the normalized $N$-qubit
state after projection is
\begin{equation}
\label{eq:primerho}
\rho'_\epsilon=
{\Pi^{\otimes N}\rho_\epsilon\Pi^{\otimes N}\over
{\rm tr}(\Pi^{\otimes N}\rho_\epsilon)}
=(1-\epsilon')M_{2^N}+\epsilon'|\Phi_{\rm cat}\rangle\langle\Phi_{\rm cat}|\;,
\end{equation}
where
\begin{equation}
|\Phi_{\rm cat}\rangle\equiv
{1\over\sqrt 2}
\Bigl(
|1\rangle\otimes\ldots\otimes|1\rangle
+|2\rangle\otimes\ldots\otimes|2\rangle
\Bigr)
\end{equation}
is the cat state for $N$ qubits and
\begin{equation}
\label{eq:eps}
\epsilon'={2\epsilon/D\over
(2/D)^N(1-\epsilon)+2\epsilon/D}\;.
\end{equation}
D{\"u}r, Cirac, and Tarrach \cite{Dur1999} and also Pittenger and Rubin
\cite{Pittenger1999} have shown that the $N$-qubit
$\epsilon$-cat state~(\ref{eq:primerho}) is nonseparable (entangled) if
and only if  $\epsilon'>1/(1+2^{N-1})$, a condition equivalent to
$\epsilon>1/(1+D^{N-1})$.  Since local projections on each qudit cannot
create entanglement, we can conclude that the $N$-qudit
$\epsilon$-cat state~(\ref{eq:mcat}) is nonseparable under the same
condition.  This establishes the upper bound~(\ref{eq:ubound}) on the
size of the separable neighborhood around the maximally mixed state.

Pittenger and Rubin \cite{Pittenger2000} have recently extended the result
of D{\"u}r, Cirac, and Tarrach \cite{Dur1999} for the $N$-qubit $\epsilon$-cat
state.  They have shown directly that the $N$-qudit
$\epsilon$-cat state~(\ref{eq:mcat}) is nonseparable if
$\epsilon>1/(1+D^{N-1})$, and they have also shown that the same condition
is a necessary and sufficient condition for entanglement when $D$ is
prime.  Their argument is akin to the correlation-coefficient argument
we give in Sect.~\ref{sec:epmax}.

\end{document}